\documentclass[12pt]{revtex4}
\usepackage{graphicx}
\usepackage{psfig}

\def\be{\begin{equation}}
\def\bea{\begin{eqnarray}}
\def\ee{\end{equation}}
\def\eea{\end{eqnarray}}

\def\eps{\varepsilon}

\def\om{\omega}

\def\dd{\mbox{d}}

\begin{document}

\title{Fermi acceleration in time-dependent rectangular billiards due to multiple passages through resonances.}

\author{ A.P. Itin$^{1,2}$ and A.I. Neishtadt$^{2,3}$}

\affiliation{
 $^{1}$Zentrum f\"{u}r optische Quantentechnologien, Universit\"{a}t
Hamburg,
 Luruper Chaussee 149, 22761 Hamburg, Germany\\
 $^{2}$Space Research Institute, Russian Academy of Sciences
  Profsoyuznaya str. 84/32, 117997 Moscow, Russia\\
 $^{3}$Department of Mathematical Sciences, Loughborough University,
 Loughborough, LE11 3TU, UK. }

\vskip 10mm

\begin{abstract}

We consider a slowly rotating rectangular billiard with moving
boundaries and use canonical perturbation theory to describe the
dynamics of a billiard particle. In the process of slow evolution
certain resonance conditions can be satisfied. Correspondingly,
phenomena of scattering on a resonance and capture into a
resonance happen in the system. These phenomena lead to
destruction of adiabatic invariance and to unlimited acceleration
of the particle.
\end{abstract}
\maketitle

{\bf \small When one slowly changes parameters of an integrable
Hamiltonian system with two degrees of freedom, two classical
actions of the unperturbed system are approximately conserved as
adiabatic invariants. However, if during slow evolution resonance
relations between frequencies of the system are satisfied,
adiabatic invariance is destroyed. The paper describes in detail
how it happens in a rectangular billiard system with moving
boundaries and how it can lead to unlimited acceleration
(so-called Fermi acceleration) of the billiard particle.}

\vskip 3mm PACS: 03.20.+i, 05.45.+b

\vskip 3mm Corresponding author: A. P. Itin. e-mail:
alx\_it@yahoo.com

\newpage

\section{Introduction}

Billiards are important models in the theory of dynamical systems
and its applications (Ref. \cite{Ar,ZS,KoTr}), especially in
optics and matter-wave optics (see, e.g., Ref.
\cite{Goos,Nockel1,Nockel2,Hent1,Hent2,Hent3,Rih,Wir,Roh,Dav,waveguide}).
Recently billiards with varying parameters became the object of
studies ( see, e.g., Ref.
\cite{Mar,INV,AI,Karlis,Petri,Edson,Dolgopyat,gel}). One of the
most interesting questions in this area of research is Fermi
acceleration (Ref. \cite{Fermi}).

In the present paper we consider acceleration of a billiard
particle due to scattering on resonances in a rectangular billiard
with slowly varying parameters. We use the methods of analysis of
these resonant phenomena designed for Hamiltonian systems with
slow and fast variables (Ref. \cite{kluwer}). The methods were
developed and the corresponding theorems were proved for smooth
systems. However, previous studies (Ref. \cite{INV,AI,waveguide})
show that they can be also applied adequately for billiard
systems, which possess discontinuities (impacts). In Ref.
\cite{INV}, acceleration of a particle due to scattering on
resonances and capture into a resonance was considered mostly for
the case of crossing periodically a single low-order resonance.
The numerical calculations revealed very slow acceleration, and a
possibility of unlimited acceleration remained an open question.
Here we describe a geometric obstacle for acceleration that
considerably slows it down, so that it is  hard to detect the
energy growth numerically. When this obstacle is not present, the
particle  accelerates without bounds fast enough to clearly see
the energy growth in numerical calculations.

\section{The model}

\begin{figure}
\includegraphics[width=90mm]{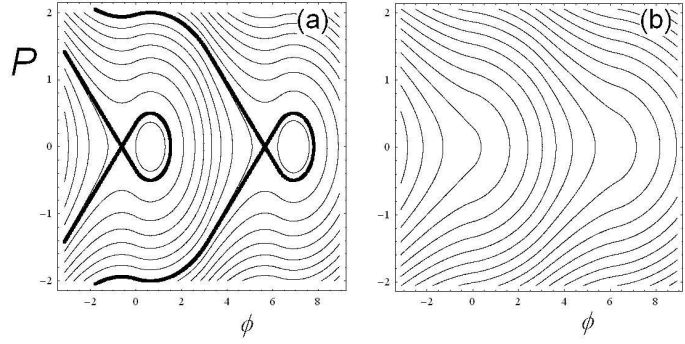}
\caption{Phase portraits of the Hamiltonian $F_0$ in (\ref{F0}).
a)  $|a|<|b|$, the phase portrait has an oscillatory domain. b)
$|a|>|b|$, there is no oscillatory domain in the phase portrait.
In both plots, $a<0$.
 } \label{Fig1}
\end{figure}

\begin{figure}
\includegraphics[width=90mm]{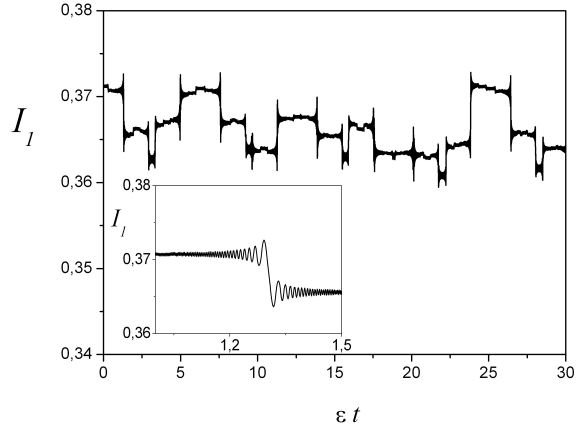}
\caption{Jumps of adiabatic invariant $I_1$. Parameters of the
system:
 $\eps=1 \cdot 10^{-4}$, $\omega=4.5 \cdot 10^{-5}$, $d_1=d_{10}(1+A_1
 \cos(\eps t))$,
$d_2=d_{20}(1+A_2 \cos(\eps t))$, where $A_1=-0.1$, $A_2=0.3$,
$d_{10}=1$, $d_{20}=1.4$.  Inset: a single jump of the adiabatic
invariant.
  } \label{JF2}
\end{figure}

Let us consider a particle in a rectangle 2D box rotating at a
constant angular velocity $\omega >0$. We assume that the rotation
is slow: $\omega \ll 1$.  The Hamiltonian of the system in the
rotating coordinate frame is

\be H=\frac{1}{2} (p_1^2+p_2^2) + \omega
(p_1q_2-p_2q_1)+U(q_1,q_2, \eps t), \label{billiard} \ee where
$p_i, q_i$ are canonically conjugate momenta and coordinates,
respectively, and $U(q_1,q_2, \eps t)$ is the hard wall potential
of the 2D box, and $\eps$ is a small positive parameter (the
time-dependence arises from motion of boundaries described below).
The second term in Eq. (\ref{billiard}) comes from switching to
the rotating frame (where billiard is irrotational): it is
$-\omega L$, where $L=-(p_1 q_2-p_2 q_1)$ is the angular momentum
of the particle about the coordinate origin \cite{Landau}. The
rotation is needed to couple two degrees of freedom, otherwise the
system would decouple on two one-dimensional billiards.

The Hamiltonian formalism outlined below is available in Ref.
\cite{INV} and is given here for self-consistency of the
presentation. Using a canonical transformation we introduce new
variables $(I_i, \phi_i), \, i=1,2$ such that

\bea
q_i &=& \left\{ -d_i+2d_i \phi_i/\pi, \quad \mbox{if} \quad 0<\phi_i \le \pi,
 \atop 3d_i-2d_i \phi_i/\pi, \quad \mbox{if} \quad \pi<\phi_i \le 2\pi,
 \right.
 \label{2} \\
p_i &=& \frac{\pi I_i}{2d_i} \mbox{sgn}(\sin\phi_i),\nonumber \eea
where $2d_i$ are lengths of the sides of the 2D box. We assume
that these lengths slowly depend on time:  $d_i=d_i(\eps t)$,
where $\eps \sim \om \ll 1$. If $d_i$ are constant, and the box
does not rotate, the new variables $(I_i,\phi_i)$ are just the
action-angle variables of the system.

In the new variables the Hamiltonian of the system is:

\bea {\cal H} = \frac{\pi^2}{8} \left( \frac{I_1^2}{d_1^2} +
\frac{I_2^2}{d_2^2} \right) - \frac{8 \omega}{\pi^2}
{\sum\limits_{k_1=1}^{\infty}}^\prime \; {\sum\limits_{k_2=1}^{\infty}}^\prime \;
 \Biggl[ \frac{(I_1d_2^2k_1+I_2d_1^2k_2)}{d_1d_2k_1^2k_2^2}
\sin (k_1\phi_1-k_2\phi_2) \Biggr.
\nonumber
\\
\Biggl. +\frac{(I_1d_2^2k_1-I_2d_1^2k_2)}{d_1d_2k_1^2k_2^2}\sin
(k_1\phi_1+k_2\phi_2) \Biggr] +\eps
\left[E_1(\phi_1)+E_2(\phi_2)\right]= H_0+ \omega H_1, \label{H}
\eea where primes denote summation over odd $k$. We use Fourier
series expansion of expressions Eq. (\ref{2}) for $p_i,q_i$ (see
Appendix). Terms $E_i$ appear because the transformation (\ref{2})
is non-autonomous; the explicit form of these terms is not
important for our discussion.

Let $m,n$ be relatively prime integers. Let us call the terms in
(\ref{H}) with $k_1=rm, k_2=rn$ and integer $r$ $(m,n)$-resonant,
if the following resonance condition is met:

\be
\frac{m I_1}{d_1^2}-\frac{n I_2}{d_2^2}=0.
\ee

Physically, the resonance condition means that frequencies of
oscillation of a particle between  pairs of opposite walls are
commensurable ($m \omega_1 = n \omega_2 $, where $\omega_j =
\frac{\pi p_j}{2 d_j}=\frac{\pi^2 I_j}{4d_j^2}, \quad j=1,2$ are
frequencies of oscillation "along" and "across" the rectangular
billiard, correspondingly). If the resonance condition is met in
the unperturbed billiard, the particle is moving along a closed
trajectory.

The resonance condition defines a corresponding resonance {\em ray
} in the plane of actions:

\be  I_2^{res} = \frac{m d_2^2 }{n d_1^2} I_1 = \alpha_{m:n} I_1,
\label{ray} \ee

where $\alpha_{m:n}$ is a slow function of time.

\section{Passage through a resonance}

\begin{figure}
\includegraphics[width=75mm]{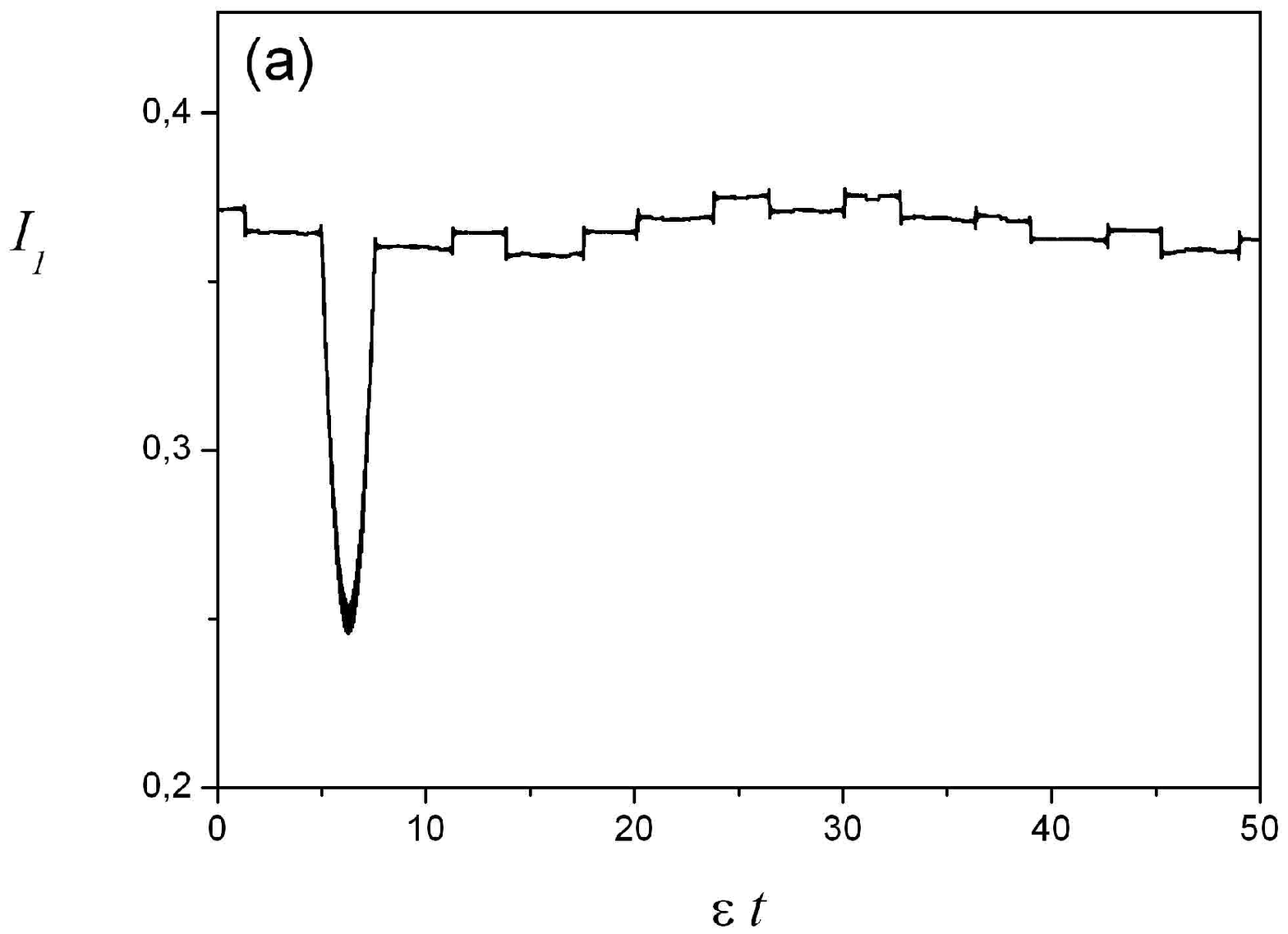}
\includegraphics[width=75mm]{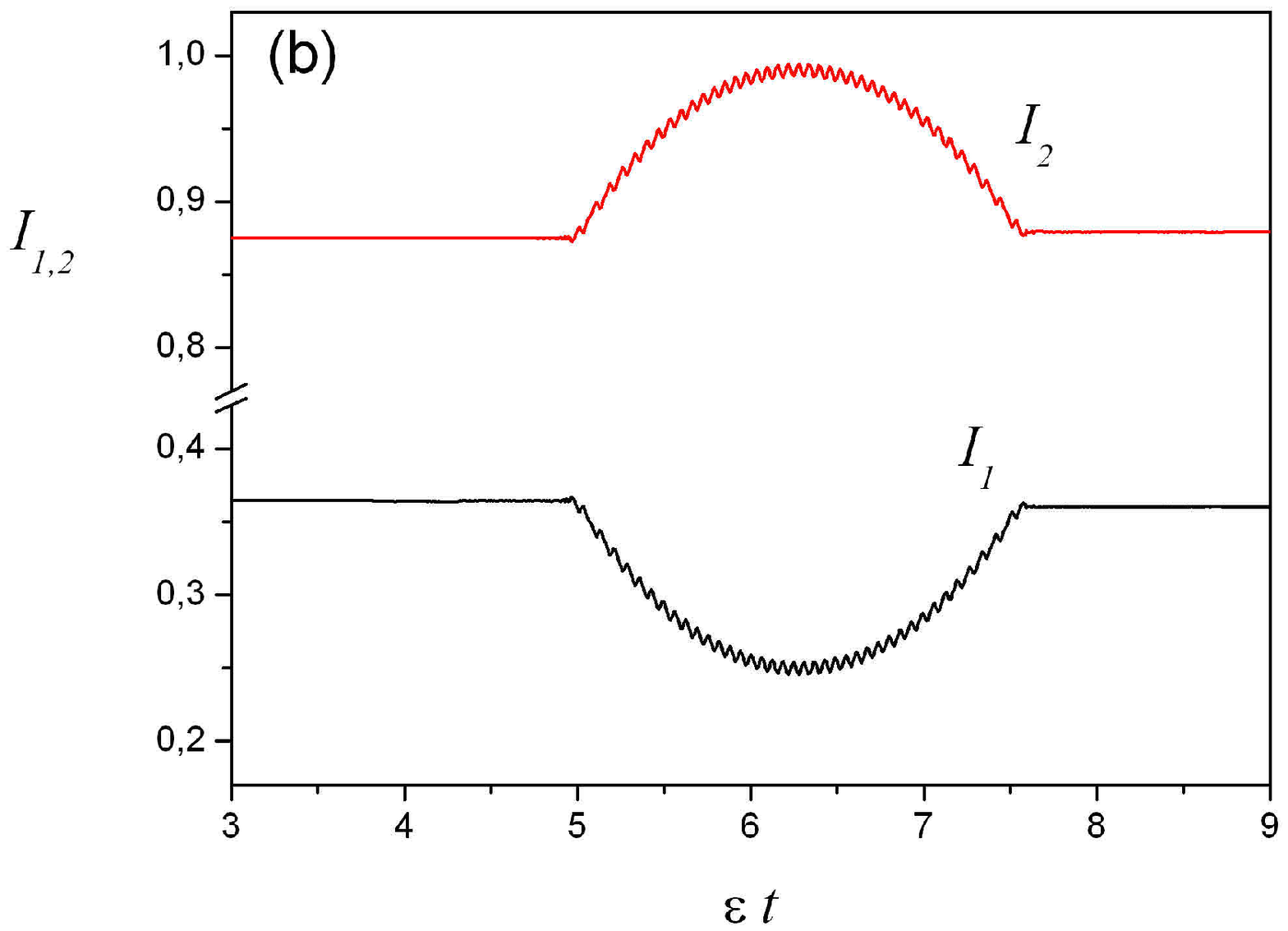}
\includegraphics[width=75mm]{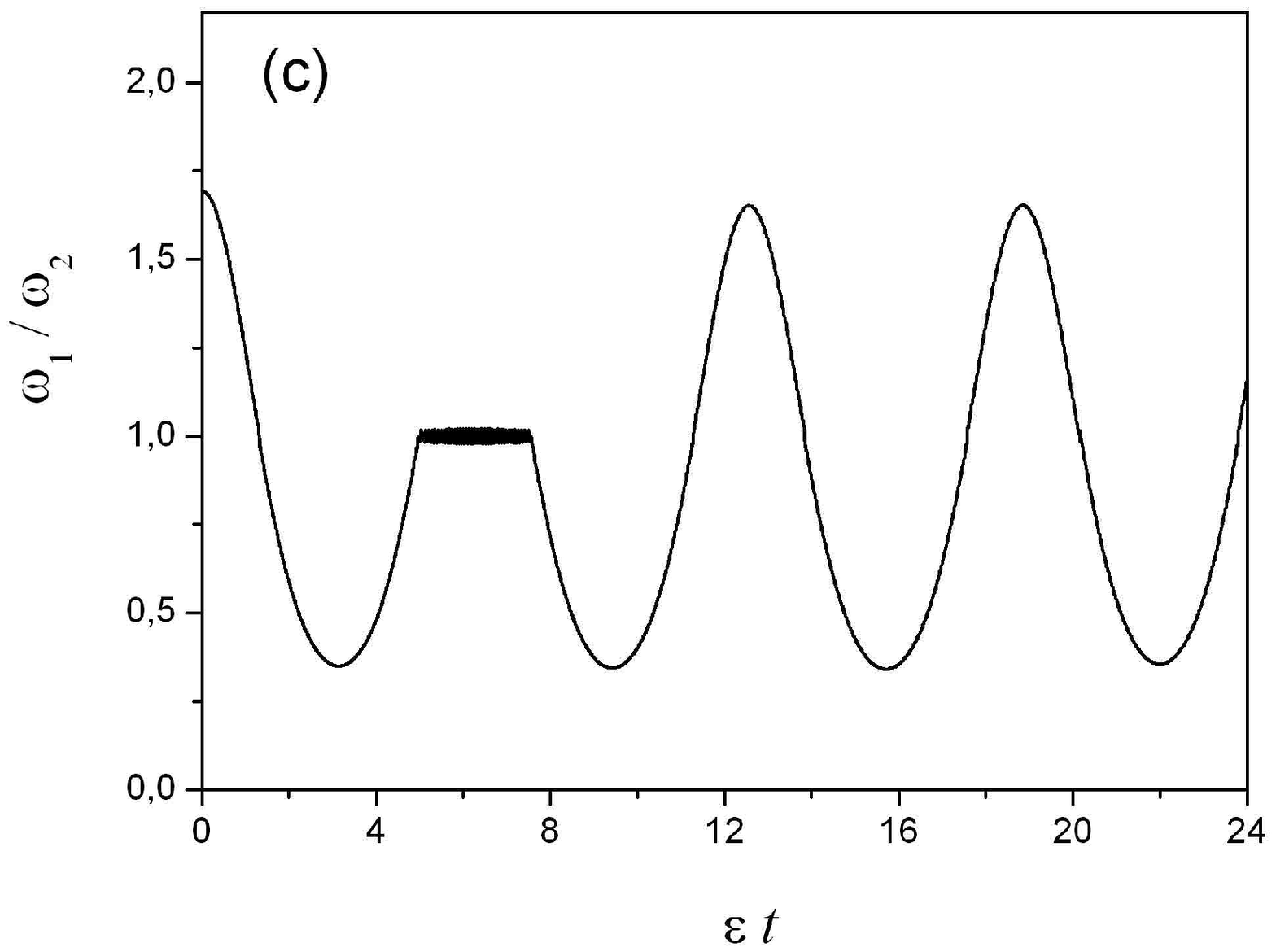}
\includegraphics[width=75mm]{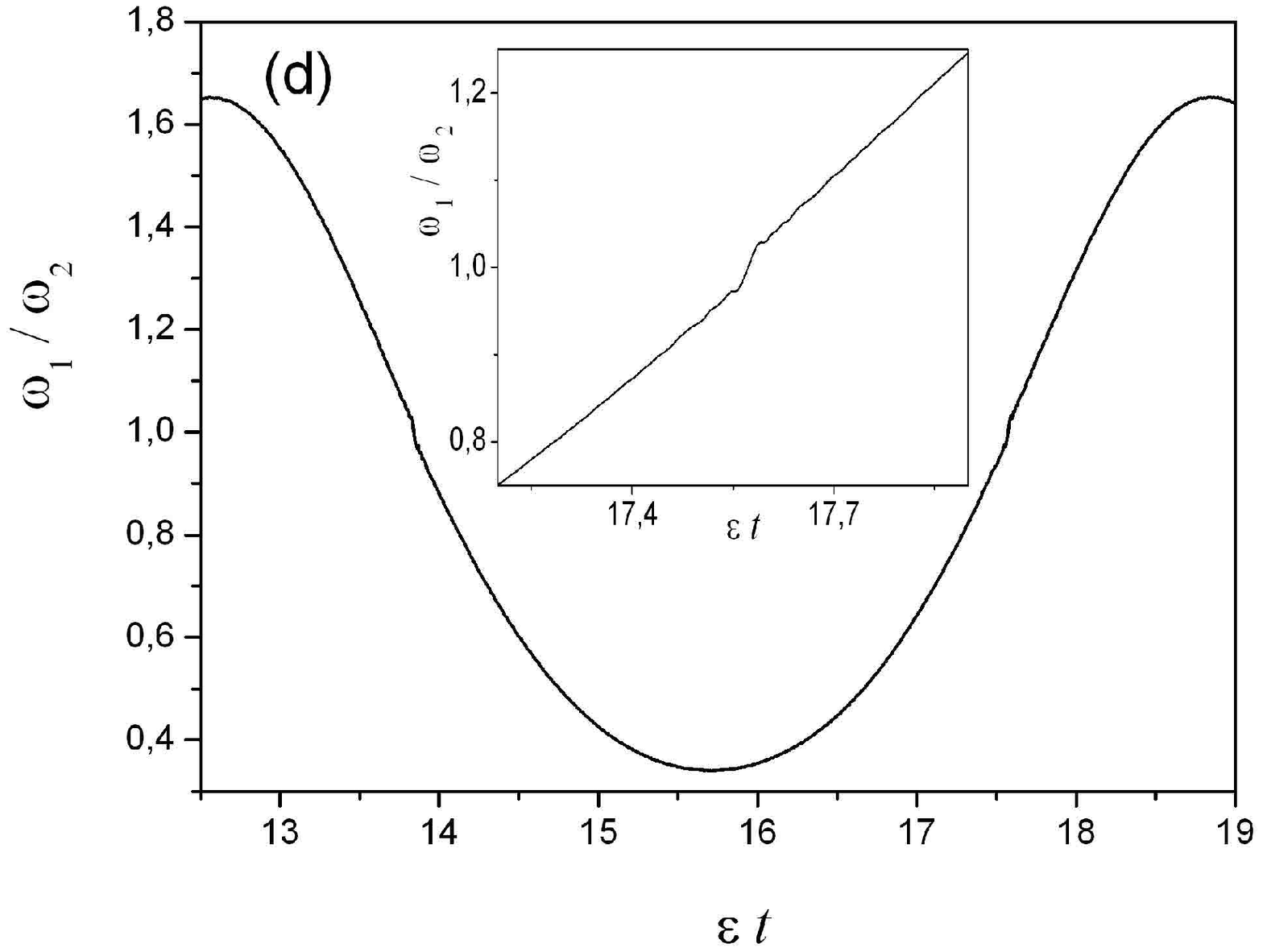}

\caption{(a) Jumps of adiabatic invariant $I_1$ (due to scattering
on a resonance) and a capture of the phase point into the 1:1
resonance. The captured point moves in the plane of actions with
the resonant ray $I_2= \alpha_{1:1} I_1$ along the resonant
trajectory $J=const$ until it escapes from the resonance. (b)
Behaviour of $I_1,I_2$ near and at the capture into the resonance
1:1 (arcs in the center of the figure correspond to the phase
point captured into the resonance)(c) Behaviour of the winding
number ($\omega_1/\omega_2$) as a function of time. For the point
captured into the resonance, this winding number stays
approximately constant and undergoes small oscillations around the
resonance value ($\omega_1/\omega_2$=1 in this particular case).
(d) The same as in panel (c), but on a different scale. Scattering
on a resonance 1:1 causes a jump in the winding number due to
jumps in $I_1,I_2$ (inset shows a single jump with even greater
resolution).
 Parameters of the system: $\eps=1 \cdot 10^{-4}$, $\omega=4.5
\cdot 10^{-5}$, $d_1=d_{10}(1+A_1 \cos(\eps t))$,
$d_2=d_{20}(1+A_2 \cos(\eps t))$, where $A_1=-0.1$, $A_2=0.28$,
$d_{10}=1$, $d_{20}=1.4$.
  \label{F3}}
\end{figure}

\begin{figure}
\includegraphics[width=100mm]{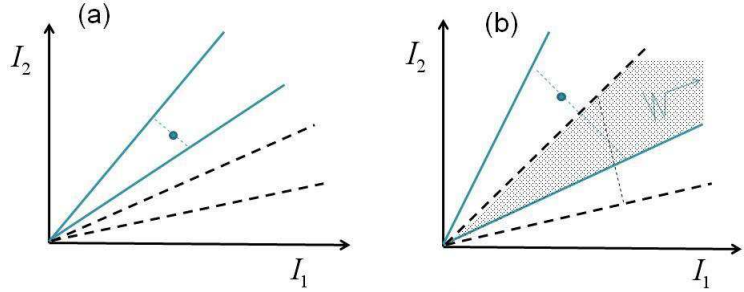}
\caption{Diffusion in the action space $(I_1,I_2)$ due to periodic
crossing of resonance lines (schematically). (a) Only one
low-order resonance is crossed. The particle can diffuse in the
action space due to kicks it receives at each passage of the
resonance ray $I_2^{res}= \alpha_{m:n}(t) I_1$ through it, but
diffusion happen only along an interval of the resonance
trajectory $J=$const within the resonance sector since all these
kicks are parallel to each other and are restricted by the
resonance sector. Solid lines shows the resonance rays $I_2 =
 \alpha_{m:n} I_1$ for two values of the slope $\alpha_{m,n}$, that
is,  $\alpha_{min}$ and $\alpha_{max}$. These two lines define the
resonance sector. Dashed lines illustrate the resonance sector of
another resonance, which our particle does not have chance to
cross. (b) Resonance sectors of several low-order resonances
overlap, correspondingly the particle consequently crosses several
resonances. At crossing of either resonance the phase point
receives a kick along the direction corresponding to the
resonance. Since these directions are not parallel ($J\equiv n I_1
+ m I_2$ is defined differently at different resonances),
unlimited diffusion is now possible. Note that each resonance is
crossed as a singe resonance, 'overlapping' means that some sector
in the action plane is sweeped by several resonance rays, but in
different times.} \label{F4}
\end{figure}

\begin{figure}
 \includegraphics[width=70mm]{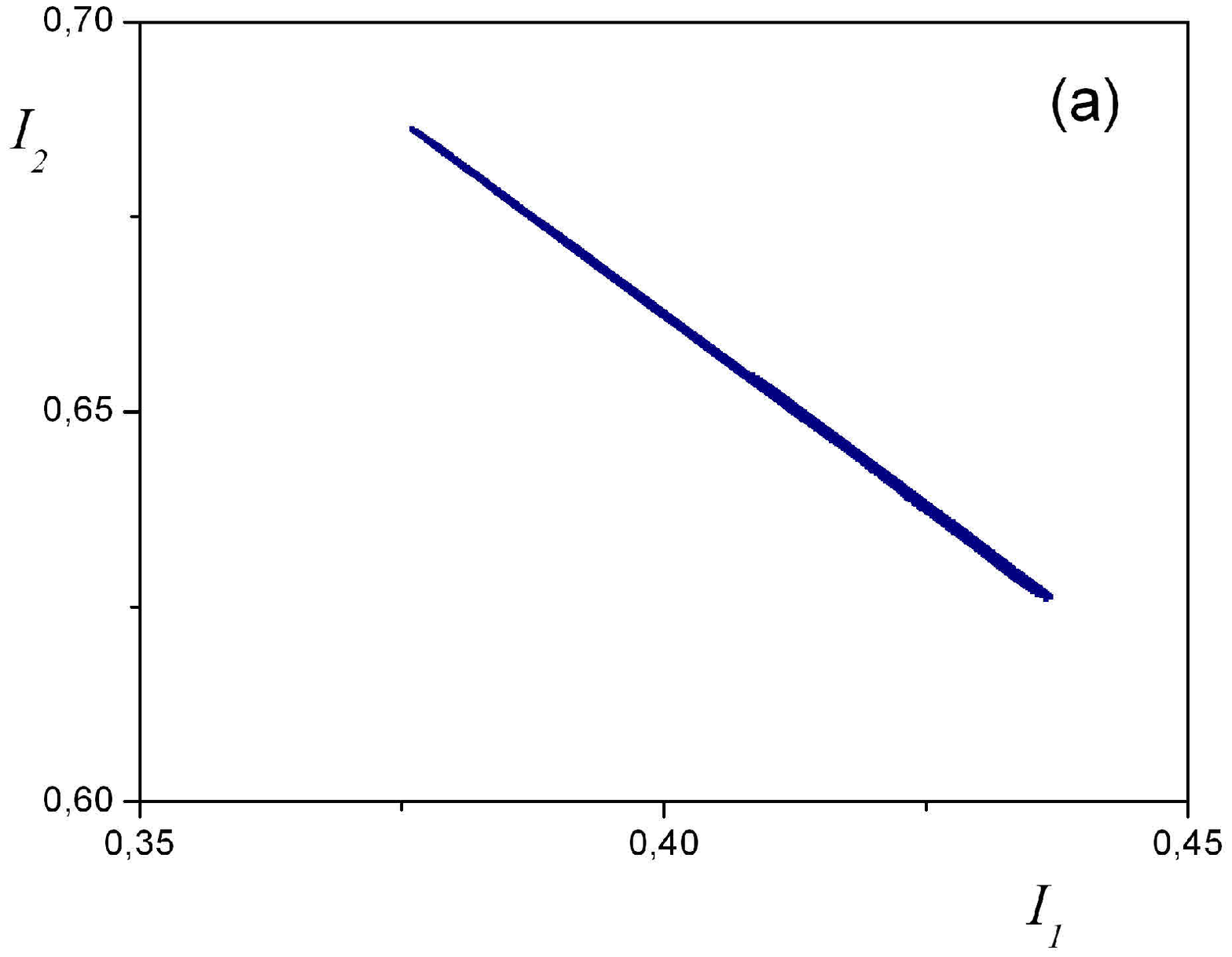}
\includegraphics[width=70mm]{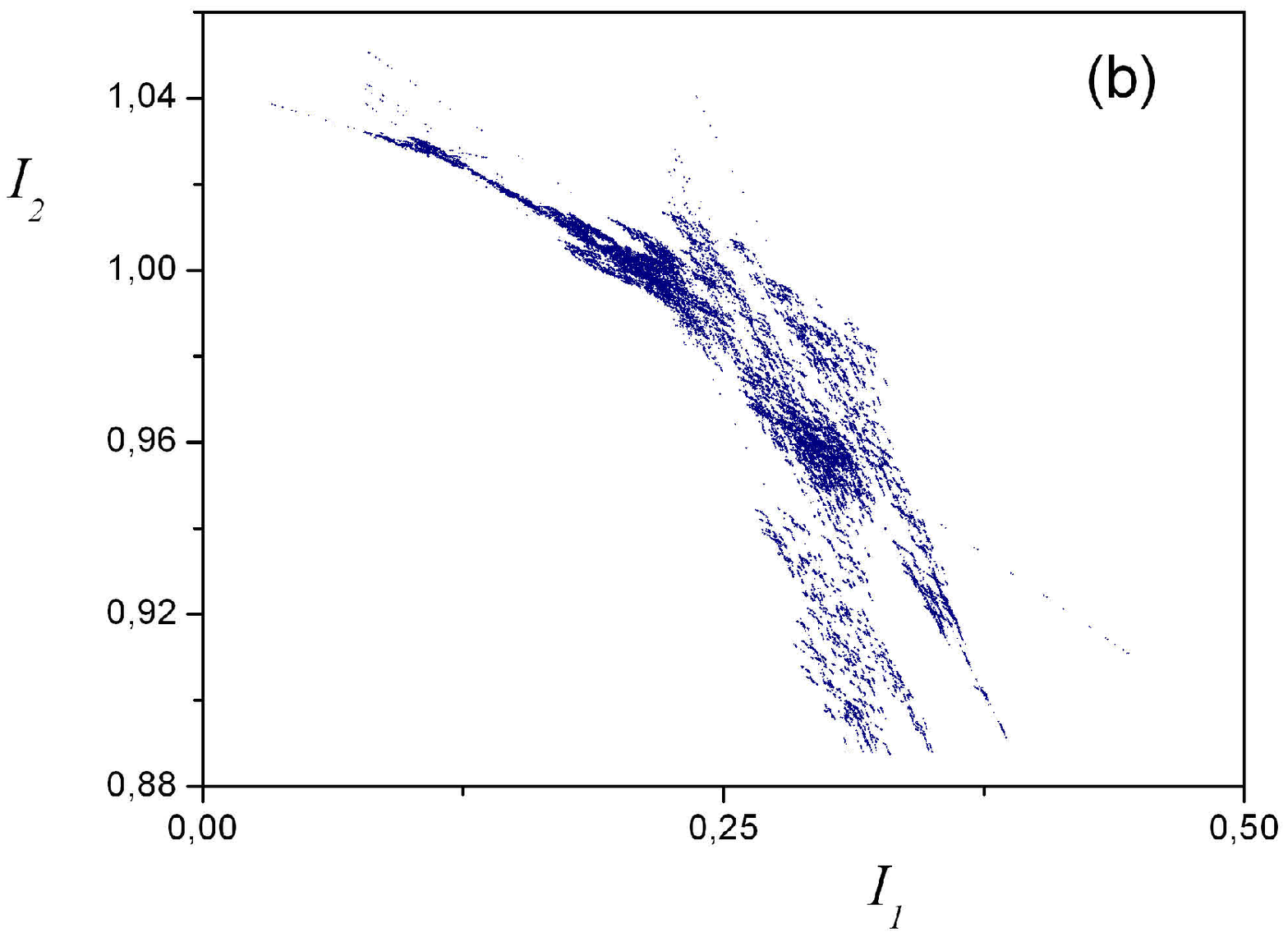}
\caption{  Diffusion in the action space. (a) Only one low-order
resonance influences dynamics of the particle, crossings of the
higher-order resonances nearby lead to slight perturbation of the
scheme outlined in Fig. 4a. The particle is almost trapped in the
resonance sector, 'transverse' diffusion due to other resonances
is very slow. Time of integration $t \approx 6 \cdot 10^7$.
Parameters are $A_1=-0.02$, $A_2=0.05$, $\omega=4.5 \cdot
10^{-5}$. (b) Several low-order resonances are crossed. Diffusion
in action space is effective due to the zigzag process outlined in
Fig. 4b. Parameters are: $A_1=-0.1$, $A_2=0.35$, $\omega=6 \cdot
10^{-5}$. Time of integration $t \approx 8 \cdot 10^7$.}
\label{F5}
\end{figure}

\begin{figure}
\includegraphics[width=70mm]{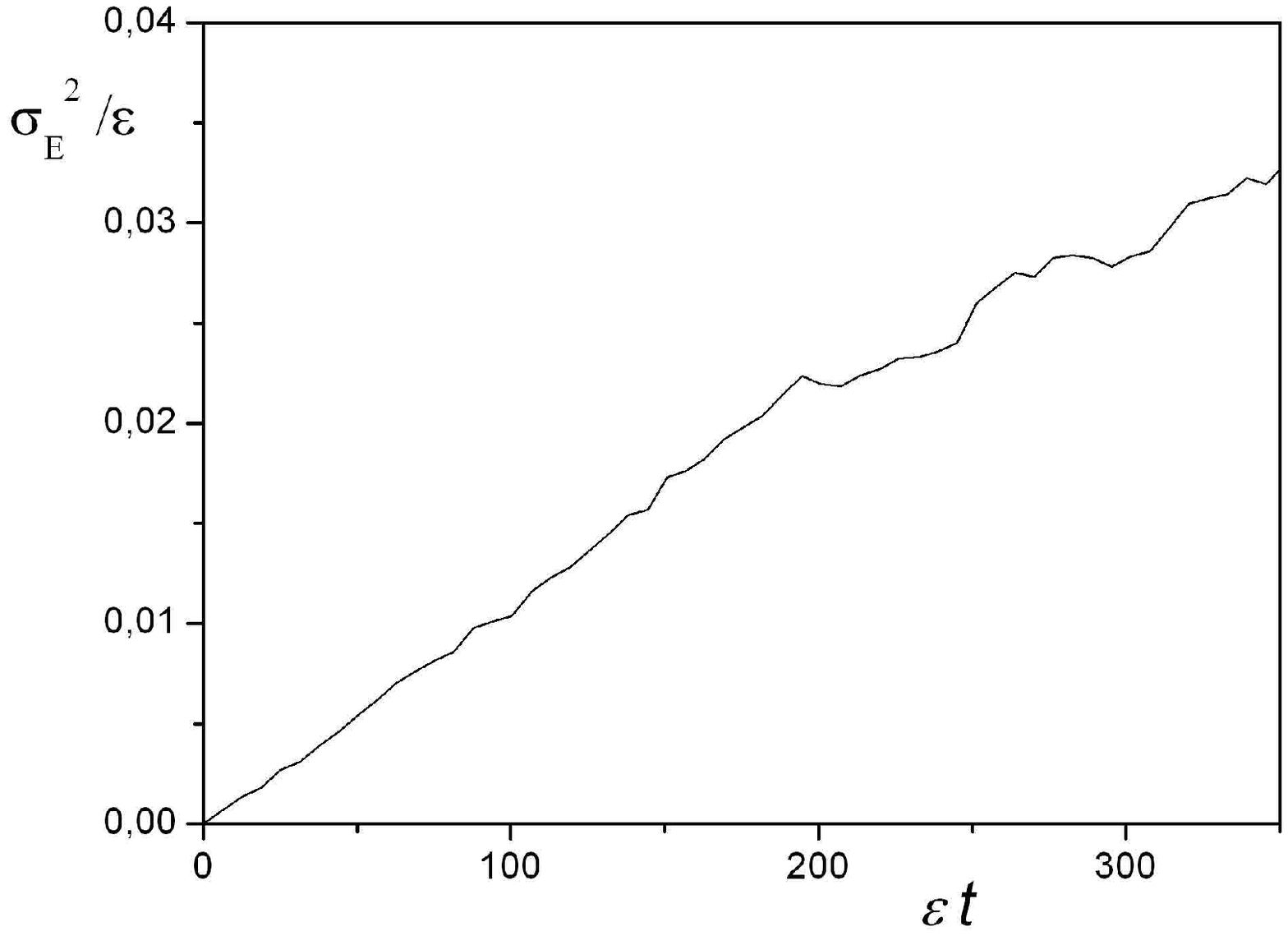}
\includegraphics[width=70mm]{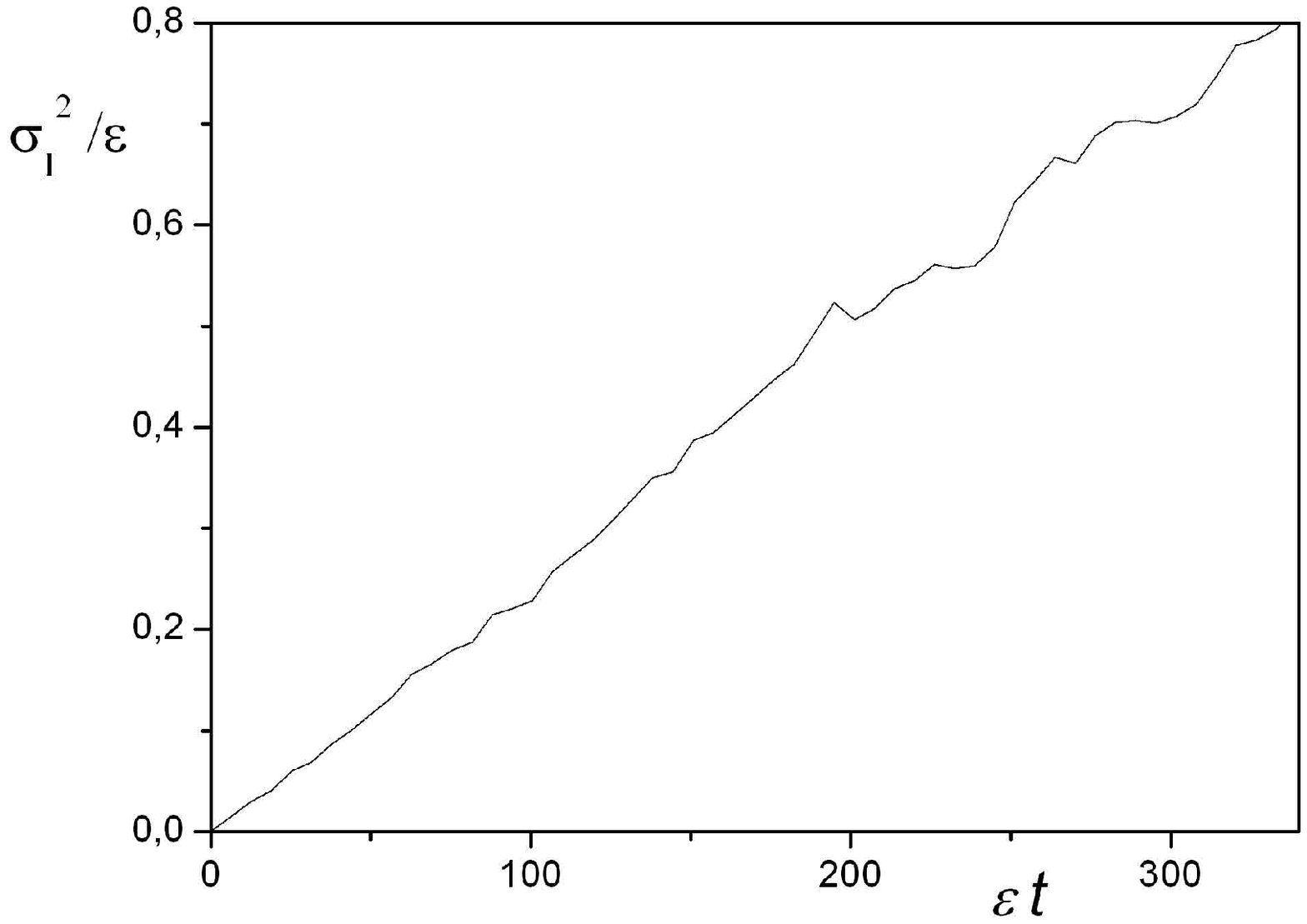}

\caption{Diffusion in energy and in action. Left: growth of
dispersion of energy of ensemble of initially isoenergetic phase
points due to crossing of a single resonance as a function of
time.  Right: growth of dispersion of action $I_1$ of the same
ensemble. In both cases, values of the magnitudes are rescaled to
$\eps$. Time of integration shown is approximately 100 slow
periods, with $\eps=0.0001$, $\omega=4.5 \cdot 10^{-5}$,
$A_1=-0.02$, $A_2=0.05$, $d_1=1$, $d_2=1.3$, number of phase
points is $N=200$.  \label{F6}}
\end{figure}

If a phase point has  such values of $I_1=I_1^{(0)}$ and
$I_2=I_2^{(0)}$ that in the plane of action it is located far from
any low-order resonance ray,  then in the process of evolution
values of $I_1$ and $I_2$ will be approximate adiabatic invariants
and the phase point will undergo small oscillations around
$(I_1^{(0)}, I_2^{(0)} )$.

However, as the time grows, the values of the lengths $d_i$ change
and the system may approach a resonance. There, actions are no
longer well-conserved, and two phenomena can happen: scattering on
resonance and capture into the resonance. They are illustrated in
Fig. 2 and Fig. 3.  Fig. 2 presents evolution of action $I_1$ in a
billiard with periodically oscillating walls. The action undergoes
small oscillations remaining almost constant until a resonance
condition is fulfilled in the system. At a resonance, it may
undergo a jump, which can be seen in detail in the inset of Fig.
2. This is scattering on a resonance. Another option for a phase
point is capture into the resonance shown in Fig.3.  As the system
approaches a resonance, the phase point can be captured in it and
continue its motion in such a way that the resonance condition
would persist. This is clearly demonstrated in Fig. 3c, where a
winding number $\omega_1/\omega_2$ is shown. Capture into the
$1:1$ resonance leads to a remarkable conservation of the winding
number during a long time (comparable to a slow period
$2\pi/\eps$). There are also several events of scattering on a
resonance which are clearly seen in Fig.3a (as jumps in the
actions similar to Fig.2), and can also be noticed in behaviour of
the winding number in Figs.3c,d. Physically, captured point moves
in such a way that its trajectory remains almost closed even
though geometry of the billiard is changing considerably.

Below we discuss these phenomena in a greater detail.

Consider the motion of a particle near a $(m,n)$-resonance.  We
make a canonical transformation $(I_1,I_2,\phi_1,\phi_2) \mapsto
(R,J,\phi,\psi)$ using the generating function
$W=(m\phi_1-n\phi_2)R- (l_1\phi_1-l_2\phi_2)J$, where $l_1, l_2$
are integers, $m l_2-n l_1=1$.  This means

\be J=n I_1 + m I_2, \quad R = l_2 I_1 + l_1 I_2  \ee

The physical meaning of this transformation is as follows. The new
phase $\phi= m \phi_1 - n \phi_2$ (the resonance phase) changes
slowly near $(m:n)-$ resonance, while the new phase $\psi = -l_1
\phi_1+ l_2 \phi_2$ changes fast. New actions $R$ and $J$ are
canonically conjugate to $\phi$ and $\psi$, respectively. We will
see that $J$ is well conserved, while the magnitude of $R$
experiences considerable dynamics in the vicinity of the
resonance. This fact is useful for the partial averaging procedure
outlined below.

The resonance condition now takes the form

\be
R=R_{res}(J,\eps t)=\frac{n l_2d_1^2+m l_1d_2^2}{m^2 d_2^2+n^2 d_1^2}J
\ee

In the new variables Hamiltonian (\ref{H}) is
${\cal{H}}={\cal{H}}_0(R,J, \eps t)+ \omega {\cal{H}}_1
(R,J,\phi,\psi, \eps t) $. The canonically conjugate pairs of
variables are $(R,\phi)$ and $(J,\psi)$. As noted above, it
follows from the form of the Hamiltonian that $\psi$ is a fast
variable, and one can average the equations of motion over it. In
the averaged system, $J$ is the integral of motion and below we
consider $J$ as a parameter. We use another canonical
transformation $(R,\phi) \mapsto (P,\phi)$ with generating
function $W_1=(P+R_{res}(J,\eps t)) \phi$ to introduce new
'action' variable $P= R-R_{res}$; the conjugate 'angle' variable
is $\phi$.  This transformation allows us to obtain an effective
Hamiltonian in a vicinity of the resonance, since the new action
$P$ measures the deviation of $R$ from its resonance value. In a
small neighborhood of the resonance where $P$ is of order
$\sqrt{\eps}$, the Hamiltonian takes the following form:

\be
{\cal{H}}=\Lambda(J, \eps t)+F_0+O(\eps), \quad F_0=\frac{1}{2}gP^2+b
\chi(\phi)+a\phi,
\label{F0}
\ee
where $\Lambda$ is ${\cal{H}}_0$ on the resonance, and

\bea
g &=& \left(\frac{\partial^2 {\cal{H}}_0}{\partial
R^2}\right)_{R= R_{res}}= \frac{\pi^2}{4} \left( \frac{k_1^2}{d_1^2} +
\frac{k_2^2}{d_2^2} \right), \quad b=-2 \frac{\omega d_1 d_2 J}{k_1 k_2
(k_1^2d_2^2 + k_2^2d_1^2)}, \quad a= \eps \frac{\partial
R_{res}}{\partial \eps t},
\nonumber\\
\\
\chi(\phi) &=& \frac{8}{\pi^2} \sum\limits_{n=1}^{\infty}\frac{1}{n^3}\sin n\phi=
\left\{\phi-\phi^2/\pi,\quad \mbox{if} \qquad 0<\phi<\pi, \atop
 \phi + \phi^2/\pi,\quad \mbox{if} \quad -\pi< \phi \le 0.
\right.
\nonumber
\eea

The dynamics near the resonance depends on the properties of the
resonance Hamiltonian $F_0$. Consider this Hamiltonian at frozen
values of $J,\eps t$. Corresponding phase portraits of the
resonance Hamiltonian on the $(\phi, P)$-plane are shown in Figure
1. If $|a|<|b|$, the phase portrait has a separatrix and an
oscillatory domain.  If $|a|>|b|$, there is no such a domain in
the phase portrait. To be definite, assume $a<0, \,b>0$. The area
$S$ of the oscillatory domain (in the case $|a|<|b|$) is a
function of $J$ and $\eps t$, and can be expressed as follows:

\be S=2\sqrt{\frac{2}{g}}
\int\limits_{\phi_1}^{\phi_2}\sqrt{F_0^s-b\chi(\phi)-a\phi} \; \dd
\phi, \label{area} \ee where $F_0^s$ is the value of $F_0$ on the
separatrix \be F_0^s =\frac{\pi}{4} \frac{(a+b)^2}{b}, \ee and
$\phi_1, \phi_2$ are the roots of the integrand (in particular,
$\phi_2$ is the coordinate of the saddle point).  The explicit
expression for the area can be found in the Appendix.

Now we take into account a slow evolution of parameter $\eps t$ in
the system. Suppose the area of the oscillation region $S$ grows
with time. In this case, additional space appears inside the
oscillatory domain. Hence, phase points can cross the separatrix
and change their mode of motion from rotational to oscillatory.

As mentioned above, this phenomenon is called a {\it capture into
a resonance}. Note that for a phase point captured into the
resonance, in the course of its motion in the phase portrait of
Hamiltonian $F_0$ with slowly changing parameters, the internal
adiabatic invariant (the area encircled by the point's orbit,
divided by $2 \pi$ ) is well conserved (see numerical calculations
in Ref. \cite{INV}).

It is important to trace the motion of the phase point in the
plane of actions $(I_1,I_2)$.

Far from the resonance, a phase point undergoes small oscillations
around its adiabatic position (a point $(I_1^{(0)},I_2^{(0)})$ in
the action plane).

As time goes, the resonance ray $I_2 = \alpha_{m:n}(\eps t) I_1$
slowly sweeps a sector (let us call it the resonance sector) on
this plane. This resonance ray is a line passing through the
origin and slowly periodically changing its slope between the
minimal and maximal values, thus sweeping the resonance sector
periodically. When the resonance ray passes through our phase
point and captures it, the captured point leaves a vicinity of the
initial adiabatic position $R = R_0 = l_2 I_1^{(0)} + l_1
I_2^{(0)}=$ const, $J=J_0=$ const in the action space $(I_1, I_2)$
and continues its motion following approximately the resonant ray
and keeping its value of $J$ constant: $R=R_{res}(J,\eps t), \,
J=J_0=$ const. Thus, it travels along  a resonant trajectory which
is a part of the line $J=J_0$ located within the resonant sector.

So, it leaves a vicinity of the initial point
$I_{1,2}=I_{1,2}^{0}$ const and continues its motion approximately
following the corresponding resonant line. Let us denote the value
of $S$ at the moment of capture as $S_*$. Somewhen later, the
value of $S$ starts decreasing with time. Approximately at the
moment when it equals the value $S_*$ the phase point previously
captured crosses the separatrix again and leaves the oscillatory
domain. This is escape from the resonance (value of $S$ at the
moment of escape is approximately equal to $S_*$ due to
conservation of the internal adiabatic invariant). If behaviour of
 the area $S(t)$ is monotonous as the slope $\alpha_{m:n}(t)$ is
 changing between $\alpha_{min}$ and  $\alpha_{max}$, e.g. it monotonously
grows as $\alpha_{m:n}$ is increasing from $\alpha_{min}$ to
$\alpha_{max}$ and then monotonously decreases as $\alpha_{m:n}$
is decreasing from $\alpha_{max}$ and $\alpha_{min}$, then the
escape from the resonance happens approximately near the same
point $(I_1^0,I_2^0)$ where the capture happened. This is the case
in our numerical examples.

Phase points that cross the resonance without being captured
undergo a jump of adiabatic invariant $R$ of order $\sqrt{\eps}$
(and, therefore, jumps of $I_{1,2}$, since $I_1=m R-l_1J$, $I_2 =
-n R+l_2 J$). As mentioned above, this phenomenon is a {\it
scattering on a resonance}. Asymptotic formula for the jump in our
case is (see also Ref. \cite{kluwer} for a general formula)

\be \Delta R = -2 \sqrt{\eps} \int \limits_{-\infty}^{\phi_*}
\frac{ \hat b \chi'(\phi) d\phi }{[2g( \hat h_* - \hat b \chi -
\hat a \phi)]^{1/2}}, \label{jump}\ee where $\hat a =a/\eps, \quad
\hat b=b/\eps$, $\phi_*$ is the phase at which resonance crossing
happen, and $ \hat h_*$ is the value of the Hamiltonian $F_0$ at
this crossing ($F_0^*$), rescaled to $\eps$: $\hat h_* =
F_0^*/\eps$. The value of $\xi \equiv$ Frac[$\hat h_*/(2 \pi \hat
a_*)$] is uniformly distributed on the interval $(0,1)$

Both capture into a resonance and scattering on a resonance are
probabilistic phenomena, and they are usual in systems with
resonance crossings (see Ref. \cite{kluwer,surfatron,PRE}).

Multiple crossings through the resonance lead to diffusion of the
adiabatic invariants due to multiple scattering on the resonance.
Note that this diffusion happen along the resonance trajectory if
only one resonance is taken into account. We discuss this question
in a more detail in the next Section.

The phase point  undergoes two crossings of the resonance per slow
period, and the mean value of the change in action is equal, in
the first approximation, to the area $S_*$ divided by $2\pi$, with
a corresponding sign depending on whether the area increases or
decreases at the moment of crossing. The mean value of the total
change in action after the two crossings is therefore zero. One
can consider  the jumps of $R$ as a random walk with timestep
equal to the half of the slow period and a quasi-random value of
the jump is obtained from Eq.(\ref{jump}) with assumption of
uniform distribution of $\xi$ (Ref.
\cite{general2,surfatron,INV}).

So, the resulting dynamics in the action space is normal
diffusion, at least far from the boundaries of the resonant sector
sweeped by the resonance ray (near the boundaries, motion of the
resonance ray slows down and the point effectively "touches" the
resonance instead of crossing it; numerically, one can see that
the point is reflected from the boundary of the sector).

The coefficient of diffusion can be found as follows. If there is
no oscillatory domain in the phase portrait of $F_0$, the mean
value of the jump (\ref{jump}) is equal to zero, and the
coefficient of diffusion is equal to the expectation value of
$(\Delta R)^2$, divided by half of the slow period, similar to
that being done in Ref. \cite{surfatron}. It can be shown that the
quasirandom variable $\xi$ is uniformly distributed on $(0,1)$ and
its values on consequent passages through the resonance are
independent. In case there is the oscillatory domain in the phase
portrait of $F_0$, the mean value of the jump $\Delta R$ is not
equal to zero, and the coefficient of diffusion can be found by
calculating the expectation value of $(\Delta R^{(1)}+\Delta
R^{(2)})^2$ from (\ref{jump}) (where $\Delta R^{(1)}$, $\Delta
R^{(2)}$ are jumps of $R$ at two consequent passages through the
resonance, with values of $\xi$ equal to $\xi_1$ and $\xi_2$,
correspondingly,  assuming the same uniform distribution of the
quasirandom variables $\xi_1$ and $\xi_2$), and dividing it by the
slow period.

Numerical calculations are presented in Figs. 2-3. For numerical
simulations we chose parameters $d_i$ harmonically varying with
time. In Figure 2 jumps of adiabatic invariant $I_1$ are shown.  A
single jump of the adiabatic invariant is shown in the inset of
Figure 2. In Figure 3a one can see the evolution of the adiabatic
invariant due to multiple scatterings on the resonance and one
capture of the phase point into the resonance. The captured point
moves along a resonant curve until it escapes from the resonance.
Figure 3b gives evolution of $I_1,I_2$ in the vicinity of capture
into resonance, Fig. 3c presents time evolution of the winding
number $\omega_1/\omega_2$.

\section{Acceleration by several resonances}

Consider what is happening if parameters $d_i(t)$ are changed
periodically in such a way that the particle repeatedly cross a
single resonance. I.e., only one low-order resonance, say n:m, is
crossed, while other resonances being crossed are of high order
and can be neglected (While there is no precise distinction
between low- and high-order resonances, in this paper we consider
as high order resonances those for which $|m|+|n|$ is bigger than
const$\cdot \eps^{-1/6}$. This is a condition that the resonance
alone does not produce any diffusion because its effect is so
small that phases of subsequent passages through this resonance
are strongly correlated, see also Ref. \cite{surfatron}). Periodic
scattering on the low-order resonance leads to diffusion in action
space. However, jumps of $I_1$ and $I_2$ at the passage through
the resonance are linearly dependent. Indeed, since $J \equiv n
I_1 + m I_2 \approx$const at the passage through the resonance,
jumps of $I_1$ and $I_2$  are dependent in such a way that in the
action plane $(I_1,I_2)$ the particle is kicked along the
corresponding line, $J=J_0=$const. Since the resonance condition
is $I_2= \alpha_{m:n}(\eps t) I_1$, one can easily see that the
resonance ray will periodically sweep a sector on the plane of
actions, and the particle will diffuse only along the part of the
line $J=n I_1+ mI_2 = $const restricted by this sector. In other
words, its diffusion in action space happens along the interval.
This is an obvious geometric obstacle to unlimited acceleration.
Of course, taking into account that near a low-order resonance
there are always high-order resonances, one gets diffusion
transverse to the interval $J=J_0$, although very slow. Now, to
let the energy of the particle to grow without bounds efficiently,
we should allow at least two low-order resonances (say, $n,m$ and
$\hat n, \hat m$), and let the corresponding resonance sectors to
overlap. Then, in the first sector diffusion happens along the
line $n I_1+ m I_2 = $const and in the second sector along the
line $\hat n I_1+ \hat m I_2 = $const. Zigzag-type acceleration is
possible:  shift of the particle in the action space due to the
first resonance is not parallel to the shift of the particle due
to the other resonance. Even though the particle is trapped within
the union of the two
 resonance sectors, it is now can
drift to infinity staying within this combined sector.
Schematically this mechanism is presented in Fig.4.

Numerical examples are given in Figs 5-6.
 In Figure 5a one can see diffusion of the phase point in the
 action space restricted by a resonance sector. The particle is effectively trapped on an interval,
 and diffusion transverse to this interval is extremely slow.  In Fig. 5b, due to
 subsequent crossings of several resonances, the particle can
 undergo unlimited diffusion. From stroboscopic shots of the
 position of the particle in action space shown in this Figure,
 peculiar properties of the motion of the particle can be recognized. There are regimes of motion where the particle
 diffuses along certain direction for a long time (when it is in a region of the action space swept by only one
 low order resonance), and there are regimes of motion, where it receives subsequent kicks in different directions
 (in the region of overlapping of resonance sectors).

 Fig. 6a demonstrates growth of dispersion in energy of an ensemble
 of phase points due to multiple scattering on a single resonance.
 It is seen that the initial growth is linear, in accordance with the theory for
 the single-resonance system. However, saturation or only extremely slow diffusion is expected
 after the ensemble of particles is redistributed within the restricted interval shown in Fig. 5a.

Similar results can be  seen in Fig. 6b,  where multiple
scattering on three low-order  resonances happens.  Since
diffusion of energy is bounded from below, we expect the total
energy of the ensemble will also start grow with time after
considerable spreading in energy happens.

It should be noted that the reasoning in this Section is of
heuristic nature. While for the partially averaged system the
formula Eq.(\ref{jump}) for the jump of action at the resonance
and the statement about conservation of $J$ are mathematically
rigorous results, the using of these results in the exact
two-frequency system is based on physically plausible reasoning.
We use numerical simulation in order to support heuristic
reasoning.

\section{Conclusion}

An important open question left in Ref. \cite{INV} concerned the
possibility of unlimited acceleration of a particle in the
billiard under consideration. It is well known (see, e.g. Ref.
\cite{ZS}) that in a similar one-dimensional problem (Ulam's model
(Ref. \cite{ZS}), motion of a particle between two periodically
moving walls) unlimited acceleration is impossible, provided that
the motion of the walls is described by smooth enough functions.
The reason is that when a particle moves fast enough with respect
to the motion of the walls, the system possesses a perpetual
adiabatic invariant (see Ref. \cite{general2}) that prevents the
particle's acceleration. In the two-dimensional model considered
in the present paper and in Ref. \cite{INV}, the adiabatic
invariance is destroyed by captures into resonances and scattering
on resonances. The hypothesis of Ref. \cite{INV} is that for a
majority of initial conditions the velocity of a particle can
reach arbitrarily large magnitudes. However,  there was not clear
numerical evidence for that in Ref. \cite{INV}.

Here we show that there are geometric obstacles to unlimited
acceleration in the case where only a single resonance influences
dynamics of the particle. The particle is then trapped in the
resonance sector. Overcoming this obstacle by modulating the
system in such a way that several low-order resonances are crossed
during a slow period of modulation, one obtains clear numerical
evidence of unlimited acceleration.

\section*{ Acknowledgements}

The work was supported in part by RFBR 09-01-00333 and
NSh-8784.2010.1 grants. The authors are grateful to
L.A.Bunimovich, D.V. Treschev, P.Schmelcher and R. de la Llave for
useful discussions. A.P.I thanks organizers of the conference
"Mathematics and physics of billiard-like systems" in Ubatuba,
Brazil, for hospitality.

\section*{Appendix}

\subsection{Fourier series}

Variables $p_i$ and $q_i$ from Eq.\ref{2} can be expressed in
terms of $(I_i,\phi_i)$ as the following Fourier series:

\bea p_i &=& {\sum\limits_{k=1}^{\infty}}^\prime \;
\frac{2I_i}{d_i k} \sin
k \phi_i \, , \\
q_i &=& -{\sum\limits_{k=1}^{\infty}}^\prime \; \frac{8 d_i}{\pi^2
k^2} \cos k \phi_i \, , \nonumber \eea where primes denote
summation over odd $k$.

Note that $I_i$ is related to $p_i$ in a very simple way: \be I_i
= \frac{2d |p_i|}{\pi}, \ee as can easily be seen from the phase
portrait of the one-dimensional billiard.  From this expression it
is easy to obtain $\frac{p_i^2}{2}=\frac{I_i^2 \pi^2}{8 d^2 },$
which give us the first term of Eq.(\ref{H}).

\subsection{Area of the oscillatory domain}

The explicit expression for $\phi_1$ in Eq.(\ref{area}) depends on
the relation between $a$ and $b$. Consider, e.g. the case
$a<(2\sqrt{2}-3)b$. One has

\be \phi_1=-\frac{\pi}{2}\frac{a+b}{b}(1+\sqrt{2}),  \quad
\phi_2=\frac{\pi}{2}\frac{a+b}{b}. \ee For the area $S$ one finds:

\be S=\sqrt{\frac{2}{g}}\frac{(a+b)^2}{2}\left( 1+\frac{3\pi}{4}
\right) \left(\frac{\pi}{b} \right)^{3/2}. \ee

\end{document}